%% file: samplepaper.tex
\documentclass[runningheads]{llncs}
\usepackage[T1]{fontenc}
\usepackage[activate={true, nocompatibility}, final, tracking=true, kerning=true, spacing=true, factor=1100, stretch=10, shrink=10]{microtype}
\usepackage{bm}
\usepackage{cite}
\usepackage{graphicx}
\input{utils}
\begin{document}
\title{Coordinate Translator for Learning Deformable Medical Image Registration}
%
%
\author{
Yihao~Liu\inst{1}
\and
Lianrui~Zuo\inst{1,2}
\and
Shuo~Han\inst{3}
\and
Yuan~Xue\inst{1}
\and
Jerry~L.~Prince\inst{1}
\and
Aaron~Carass\inst{1}
}
%
%
\authorrunning{Y. Liu \etal}
%
%
%
\institute{Department of Electrical and Computer Engineering, \\
Johns Hopkins University, Baltimore, MD 21218 USA
\and
Laboratory of Behavioral Neuroscience, National Institute on Aging, \\
National Institute of Health, Baltimore, MD 20892 USA
\and
Department of Biomedical Engineering, \\
Johns Hopkins University, Baltimore, MD 21218 USA
}
\maketitle              
\begin{abstract}
The majority of deep learning (DL) based deformable image
registration methods use convolutional neural networks~(CNNs) to
estimate displacement fields from pairs
of moving and fixed images. This, however, requires the convolutional
kernels in the CNN to not only extract intensity features from the inputs but also
understand image coordinate systems. We argue that the latter task is challenging
for traditional CNNs, limiting their performance in registration tasks.
To tackle this problem, we first introduce Coordinate Translator,
a differentiable module that identifies matched features between the fixed
and moving image and outputs their coordinate correspondences without the
need for training. It unloads the burden of understanding image coordinate systems for CNNs, allowing them to focus on feature extraction.
We then propose a novel deformable registration network, \texttt{im2grid}, that uses multiple Coordinate Translator's with the hierarchical features extracted from a CNN encoder and outputs a deformation field in a coarse-to-fine fashion.
We compared \texttt{im2grid} with the state-of-the-art DL and non-DL methods for unsupervised 3D magnetic resonance image registration. Our
experiments show that \texttt{im2grid} outperforms these methods both qualitatively and quantitatively.
\end{abstract}
\section{Introduction}
Deformable registration is of fundamental importance in medical
image analysis. Given a pair of images, one fixed and one moving, 
deformable registration warps the moving image by optimizing
the parameters of a nonlinear transformation so that the underlying anatomies of the two images are aligned according to an image dissimilarity function~\cite{davatzikos1997spatial,ferrant2000registration,rueckert1999nonrigid,thirion1998image,vercauteren2009diffeomorphic}.
Recent deep learning~(DL) methods use convolutional neural networks~(CNNs) whose parameters are optimized during training;
at test time, a dense displacement field that represents the deformable transformation
is generated in a single forward pass.

Although CNN-based methods for segmentation and classification are better than traditional methods in both speed and accuracy, DL-based deformable registration methods are faster but usually not more accurate~\cite{balakrishnan2019voxelmorph,fan2019birnet,vos2017end,yang2017quicksilver,chen2021vit}.
Using a CNN for registration requires learning coordinate correspondences between image pairs, which has been thought to be fundamentally different from other CNN applications because it involves both extracting and matching features\cite{dosovitskiy2015flownetS,ilg2017flownetS}.
However, the majority of existing works simply rely on CNNs to implicitly learn the displacement between the fixed and moving images~\cite{balakrishnan2019voxelmorph,fan2019birnet,vos2017end}.

Registration involves both feature extraction and feature matching,
but to produce a displacement field, matched features need to be translated
to coordinate correspondences.
We argue that using convolutional kernels for the latter two tasks is not optimal.
To tackle this problem, we introduce Coordinate Translator, a differentiable module that
matches features between the fixed and moving images and identifies feature matches as
precise coordinate correspondences without the need for training. 
The proposed registration network, named \texttt{im2grid}, uses multiple Coordinate Translator's with
multi-scale feature maps. These produce multi-scale sampling grids representing
coordinate correspondences, which are then composed in a coarse-to-fine manner
to warp the moving image. \texttt{im2grid} explicitly handles the task of matching features and establishing coordinate correspondence using Coordinate Translator's, leaving only feature extraction to our CNN encoder.

Throughout this paper, we use unsupervised 3D magnetic resonance~(MR)
image registration as our example task and demonstrate that the proposed
method outperforms the state-of-the-art methods in terms of registration accuracy. We think it is important to note that because
producing a coordinate location is such a common task in both medical
image analysis and computer vision, the proposed
method can be impactful on a board range of applications.

\section{Related Works}
Traditional registration methods solve an
optimization problem for every pair of fixed, $I_{f}$, and moving ,$I_{m}$, images.
Let $\phi$ denote a transformation and let the best transformation
$\hat{\phi}$ be found from
\begin{equation}
    \hat{\phi} = \argmin_{\phi}L_{\text{sim}}(I_f, I_m\circ\phi) + \lambda L_{\text{smooth}}(\phi),
    \label{e:optimization}
\end{equation}
where $I_{m}\circ\phi$ yields the warped image $I_{w}$. 
The first term focuses on the similarity between $I_{f}$ and $I_{m}\circ\phi$\, whereas
the second term---weighted by the hyper-parameter $\lambda$---regularizes
$\phi$. The choice of $L_{\text{sim}}$ is application-specific.
Popular methods using this
framework include spline-based free-form
deformable models~\cite{rueckert1999nonrigid}, elastic warping
methods~\cite{davatzikos1997spatial,klein2009elastix},
biomechanical models~\cite{ferrant2000registration},
and Demons~\cite{thirion1998image,vercauteren2009diffeomorphic}.
Alternatively, learning-based methods have also been used to estimate
the transformation parameters~\cite{chou20132d,gutierrez2016learning}. 

Recently, deep learning~(DL) methods, especially CNNs, have been used for solving deformable registration problems. In these methods, $\phi$ is typically represented as a map of
displacement vectors that specify the voxel-level spatial offsets
between $I_{f}$ and $I_{m}$;  the CNN is trained
to output $\phi$ with or without supervision~\cite{balakrishnan2019voxelmorph,cao2017deformable,fan2019birnet,han2021deformableS,vos2017end}.
In the unsupervised setting, the displacement field is converted to a sampling grid and the warped image is produced
by using a grid sampler~\cite{jaderberg2015spatial} with the moving
image and the sampling grid as input. The grid sampler performs
differentiable sampling of an image~(or a multi-channel feature map)
using a sampling grid; it allows the dissimilarity loss computed
between the warped and fixed images to be back-propagated so the CNN can be trained end-to-end.
In past work, \cite{balakrishnan2019voxelmorph} used a U-shaped network to output
the dense displacement; \cite{de2019deep,vos2017end} used an encoder network to
produce a sparse map of control points and generated the dense displacement field
by interpolation; and \cite{chen2021vit} replaced the bottleneck of a U-Net~\cite{ronneberger2015u} with a transformer structure~\cite{vaswani2017attention}.
Several deep learning methods also demonstrate the possibility of using a
velocity-based transformation representation to enforce
a diffeomorphism~\cite{dalca2018unsupervised,yang2017quicksilver}. 

Our method represents the transformation using a sampling grid $G$,
which can be directly used by the grid sampler.
For $N$-dimensional images ($N=3$ in this paper), $G$ is represented by an $N$-channel map. 
Specifically, for a voxel coordinate $\bm{x} \in \mathbb{D}^N$~(where $\mathbb{D}^{N}$ contains all the voxel coordinates in $I_{f}$),  
$G(\bm{x})$ should ideally hold a coordinate such that the two 
values $I_{f}( \bm{x} )$ and $I_{m}( G ( \bm{x} ) )$ represent 
the same anatomy.  Note that the displacement
field representation commonly used by other methods can be found as $G-G_{I}$,
where $G_{I}$ is the identity grid $G_{I}( \bm{x} ) = \bm{x}$.

\section{Method}
For the image pair $I_{f}$ and $I_{m}$,
the proposed method produces a sampling grid $G_{0}$ that can be used by the grid sampler
to warp $I_{m}$ to match $I_{f}$. Similar to previous DL methods, we use a CNN encoder to extract
multi-level feature maps from $I_{f}$ and $I_{m}$. Instead of directly producing a single displacement field from the CNN, $G_{0}$ is the composition of
multi-level sampling grids, generated from the multi-level feature maps with the proposed Coordinate Translator's.

\subsection{Coordinate Translator}
Let $F$ and $M$ denote the multi-channel feature maps that are individually extracted from $I_f$ and $I_m$, respectively. 
The goal of a Coordinate Translator is to take as input both $F$ and $M$, and produce a sampling grid $G$ that
aligns $M$ interpolated at coordinate $G(\bm{x})$ with $F(\bm{x})$ for all $\bm{x} \in \mathbb{D}^{N}$. 

As the first step, for every $\bm{x}$, cross-correlation is calculated between
$F(\bm{x})$ and $M(\bm{c}_{i})$ along
the feature dimension, where $\bm{c}_{i} \in \mathbb{D}^{N}$ for $i\in[1, K]$ are a set of candidate coordinates.
The results are a $K$-element vector of matching scores between $F(\bm{x})$ and every $M(\bm{c}_{i})$:

\begin{equation}
    \text{Matching Score}(\bm{x}) = \left( F(\bm{x})^{T} M(\bm{c}_{1}), \ldots, F(\bm{x})^{T} M(\bm{c}_{K}) \right).
\end{equation}

The choice of $\bm{c}_{i}$'s determines the search region for
the match. For example, defining $\bm{c}_{i}$ to be every coordinates in $\mathbb{D}^{N}$ 
will compare $F(\bm{x})$ against every location in $M$; these matches can also be restricted within the $3\times3\times3$ neighborhood of $\bm{x}$. We outline our choices of $\bm{c}_{i}$'s in Sec.~\ref{s:expt}.
The matching scores are normalized using a softmax function to produce a matching probability $p_{i}$,

\begin{equation}
    p_{i} = \frac{\exp \left( F(\bm{x})^{T} M(\bm{c}_{i}) \right)}{\sum_{j} \exp \left( F(\bm{x})^{T} M(\bm{c}_{j} ) \right)} \qquad\mbox{for every } \bm{c}_{i}.
\end{equation}

We interpret the matching probabilities as the strength of attraction between $F(\bm{x})$ and the $M(\bm{c}_{i})$'s.
Importantly, we can calculate a weighted sum of $\bm{c}_{i}$'s to produce a coordinate $\bm{x}' \in \mathbb{R}^{N}$, i.e., $\bm{x}' = \sum_{i=1}^{K}p_{i} \cdot \bm{c}_{i}$,
%
%
%
%
%
which represents the correspondence of $\bm{x}$ in the moving image $I_{m}$. This is conceptually similar to
the combined force in the Demons algorithm~\cite{thirion1998image}. For every $\bm{x} \in \mathbb{D}^{N}$ the corresponding $\bm{x'}$ forms the Coordinate Translator output, $G$.

\begin{figure}[!tb]
    \centering
    \includegraphics[width=1\textwidth]{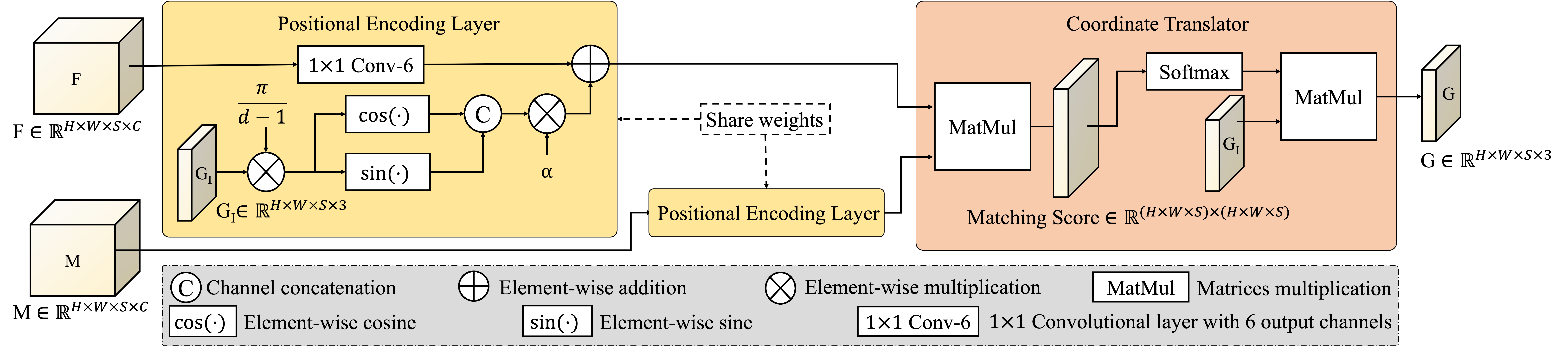}
    \caption{Structure of the proposed positional encoding layer
    and Coordinate Translator.}
    \label{f:coordinate_translator}
\end{figure}

Coordinate Translator can be efficiently implemented as the
Scaled Dot-Product Attention introduced in the Transformer~\cite{vaswani2017attention}
using matrix operations. For
3D images with spatial dimension $H\times W\times S$ and $C$ feature channels, we reshape $F$ and $M$ to
$\mathbb{R}^{(H\times W\times S)\times C}$ and
the identity grid $G_{I}$ to
$\mathbb{R}^{(H\times W\times S)\times 3}$. Thus Coordinate Translator
with $\left\{ \bm{c}_{1},\ldots \bm{c}_{K}\right\} = \mathbb{D}^{N}$ can be readily computed from,
\begin{equation}
    \text{Coordinate Translator}(F, M) = \text{Softmax}(F M^{T})G_{I},
\end{equation}
with the softmax operating on the rows of $F M^{T}$.

\subsubsection{Positional encoding layer}

In learning transformations, it is a common practice to initialize from~(or close to) an identity
transformation~\cite{jaderberg2015spatial,balakrishnan2019voxelmorph,chen2021vit}.
As shown in Fig.~\ref{f:coordinate_translator}, we propose a
positional encoding layer that combines position information with $F$ and $M$ such that the
initial output of Coordinate Translator is an identity grid.
Inside a positional encoding layer, for every $\bm{x} = \left(x_1, \cdots, x_N \right)$ with $x_{i}$'s on an integer grid ($x_i \in \{0, \ldots, d_{i} - 1\}$), we add a positional embedding~(PE),
%
\begin{equation*}
    \text{PE}(\bm{x}) = \left(\cos \frac{x_1\pi}{d_{1} - 1}, \sin \frac{x_1\pi}{d_{1} - 1}, \cdots,  \cos \frac{x_N\pi}{d_{N} - 1}, \sin \frac{x_N\pi}{d_{N} - 1} \right),
\end{equation*}
to the input feature map, where $d_{i}$ is the pixel dimension along the $i^{th}$ axis. Trigonometric identities give the cross-correlation of PEs at $\bm{x}_1$ and $\bm{x}_2$ as
\begin{equation*}
    \text{PE}(\bm{x}_{1})^{T} \text{PE}(\bm{x}_{2}) = \sum_{i = 1}^{N} \cos \left( \frac{\Delta{} x_{i} \pi}{d_{i} - 1} \right),
\end{equation*}
where $\Delta{} x_{i}$ is the difference in the $i^{\mbox{\tiny{~th}}}$
components of $\bm{x}_{1}$ and $\bm{x}_{2}$. This
has maximum value when $\bm{x}_1 = \bm{x}_2$ and decreases with the $L_{1}$ distance between the two coordinates. We initialize the convolutional layer to have zero weights and bias
and the learnable parameter $\alpha = 1$~(see Fig.~\ref{f:coordinate_translator}) such that only
the PEs are considered by Coordinate Translator at the beginning of training. As a result,
among all $\bm{c}_{i} \in \mathbb{D}^{N}$, $M(\bm{x})$ will have the highest matching score with $F(\bm{x})$, thus producing $G_{I}$ as the initial output.
Coordinate Translator also benefits from incorporating the position information as it allows the relative distance between $\bm{c}_{i}$ and $\bm{x}$ to contribute to the matching scores, similar as the positional embedding in the Transformer~\cite{vaswani2017attention}.
\begin{figure}[!t]
    \centering
    \includegraphics[width=1\textwidth]{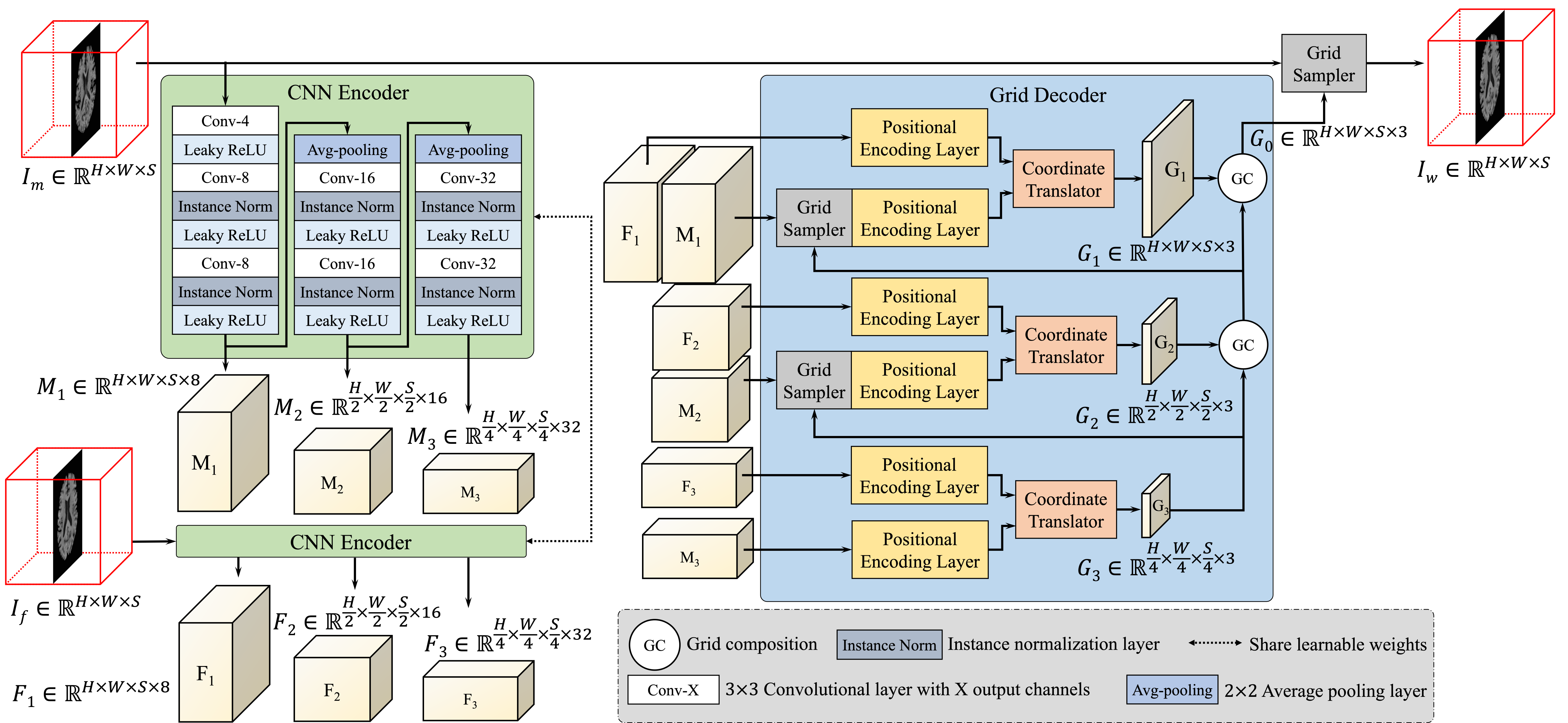}
    \caption{Example of the proposed \texttt{im2grid} network structure with a 3-level CNN encoder. The grid composition operation can be implemented using the grid sampler with two grids as input.}
    \label{f:network}
\end{figure}

\subsection{\texttt{im2grid} Network Architecture}
\begin{figure}[!t]
    \centering
    \includegraphics[width=1\textwidth]{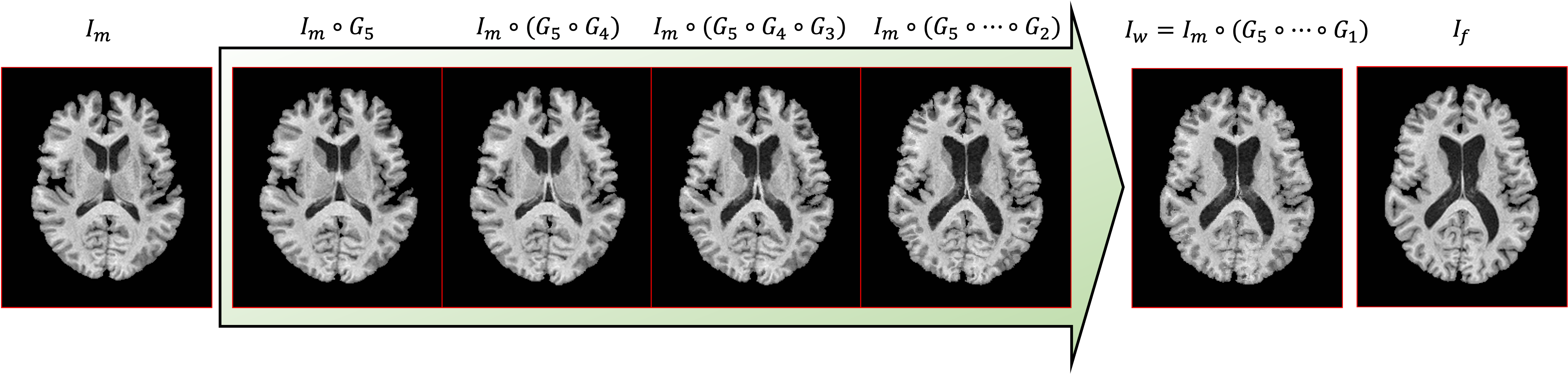}
    \caption{Visualization of the multi-scale sampling grids by
    sequentially applying finer grids to the moving image. Here
    we used a five-level CNN encoder and $G_{5}, \ldots, G_{1}$ are coarse to fine sampling grids produced from the
    five-level feature maps.}
    \label{f:visualize_sequence}
\end{figure}

The proposed \texttt{im2grid} network is shown in Fig.~\ref{f:network}.
Similar to previous methods, \texttt{im2grid} produces a sampling grid to warp $I_{m}$ to $I_{w}$. Our CNN encoder uses multiple pooling layers to extract hierarchical features
from the intensity images. In the context of intra-modal
registration, it is used as a Siamese network that processes
$I_{f}$ and $I_{m}$ separately. For clarity, Fig.~\ref{f:network} only shows a three level \texttt{im2grid} model with three level feature maps $F_{1}/F_{2}/F_{3}$ and $M_{1}/M_{2}/M_{3}$ for $I_{f}$ and $I_{m}$, respectively. In our experiment, we used a five level structure.
Our grid decoder uses the common coarse-to-fine strategy in registration. Firstly, coarse features $F_{3}$ and $M_{3}$ are matched and translated to a coarse sampling grid $G_{3}$ using
a Coordinate Translator. Because of the pooling layers, this can be interpreted as matching downsampled versions of $I_{f}$ and $I_{m}$, producing a coarse displacement field.
$G_{3}$ is then used to warp $M_{2}$, resolving
the coarse deformation between $M_{2}$ and $F_{2}$ so that the Coordinate Translator at the second level can capture
more detailed displacements with a smaller search region. Similarly, $M_{1}$ is warped by the composed transformation of $G_{3}$ and
$G_{2}$ and finally the moving image is warped by the composition of the transformations from all levels.
A visualization of a five-level version of our multi-scale sampling grids is provided in Fig.~\ref{f:visualize_sequence}. 
In contrast to previous methods that use CNNs to directly output displacements,
our CNN encoder only needs to extract
similar features for corresponding
anatomies in $I_{f}$ and $I_{m}$ and the exact coordinate
correspondences are obtained by Coordinate Translator's.
Because our CNN encoder processes $I_f$ and $I_m$ separately, it is guaranteed that our CNN encoder only performs feature extraction.

The proposed network is trained using the mean squared difference
between $I_{f}$ and $I_{w} (= I_{m} \circ \phi)$ and a smoothness loss that regularizes the spatial variations of the $G$'s at every level,
\begin{equation}
    \mathcal{L} = \frac{1}{\left|\mathbb{D}^{N}\right|} \sum_{\bm{x} \in \mathbb{D}^{N}} \left({I_{f}( \bm{x} ) - I_{w}( \bm{x} )} \right)^{2} + \lambda \sum_{i} \sum_{\bm{x} \in \mathbb{D}^{N}} || \nabla(G_{i}(\bm{x}) - G_{I}( \bm{x} ))||^{2},
\end{equation}
where $\left|\mathbb{D}^{N}\right|$ is the cardinality of $\mathbb{D}^{N}$ and all $G_{i}$'s and $G_{I}$ are normalized to $[-1,1]$.

\section{Experiments}
\label{s:expt}
\subsubsection{Datasets}
We used the publicly available \texttt{OASIS3}~\cite{lamontagne2019oasis} and \texttt{IXI}~\cite{ixi} datasets in our experiments.
$200$, $40$, and $100$ T1-weighted~(T1w) MR images of the human brain from the \texttt{OASIS3} dataset were used for training, validation, and testing, respectively.
During training, two scans were randomly selected as $I_{f}$ and $I_{m}$,
while validation and testing used $20$ and $50$ pre-assigned image pairs, respectively.
For the \texttt{IXI} dataset, we used $200$ scans for training, $20$ and $40$ pairs for validation and testing, respectively.
All scans underwent N4 inhomogeneity correction~\cite{tustison2010n4itk}, and were rigidly registered to MNI space~\cite{fonov-unbiased-2009} with $1$~mm$^3$~(for IXI) or $0.8$~mm$^3$~(for OASIS3) isotropic resolution. 
A white matter peak normalization~\cite{reinhold2019evaluatingS} was applied to standardize the MR intensity scale.

\subsubsection{Evaluation Metrics}
First, we calculated the Dice similarity
coefficient~(DSC) between segmentation
labels of $I_{f}$ and the warped labels of $I_{m}$.
An accurate transformation should align the structures of
the fixed and moving images and produces a high DSC.
We obtained a whole brain segmentation for the fixed and moving images using 
SLANT~\cite{huo20193d} and
combined the SLANT labels ($133$ labels) to TOADS labels ($9$ labels)~\cite{bazin2007topology}.
The warped labels were produced by applying each methods deformation field to the moving image labels.
Second, we measured the regularity of the transformations by computing the determinant of the Jacobian matrix, which should be globally positive for a diffeomorphic transformation.

\subsubsection{Implementation Details} Our method was implemented using
PyTorch and trained using the Adam optimizer with a learning rate of
$3\times10^{-4}$, a weight decay of $1\times10^{-9}$, and a batch
size of $1$. Random flipping of the input volumes along the three axes
were used as data augmentation. We used a five-level structure
and tested different choices of $\bm{c}_{i}$'s for each Coordinate Translator. We found that
given the hierarchical structure, a small search region at each level
is sufficient to capture displacements presented in our data.
Therefore, we implemented two versions of our
method: 1)~\texttt{im2grid} which used a $3\times3$
search window in the axial plane for producing $G_{1}$ and a
$3\times3\times3$ search window at other levels;
and 2)~\texttt{im2grid-Lite} which is identical to \texttt{im2grid}
except that the finest grid $G_{1}$ is not used. 

\begin{figure}[!t]
    \centering
    \includegraphics[width=1\textwidth]{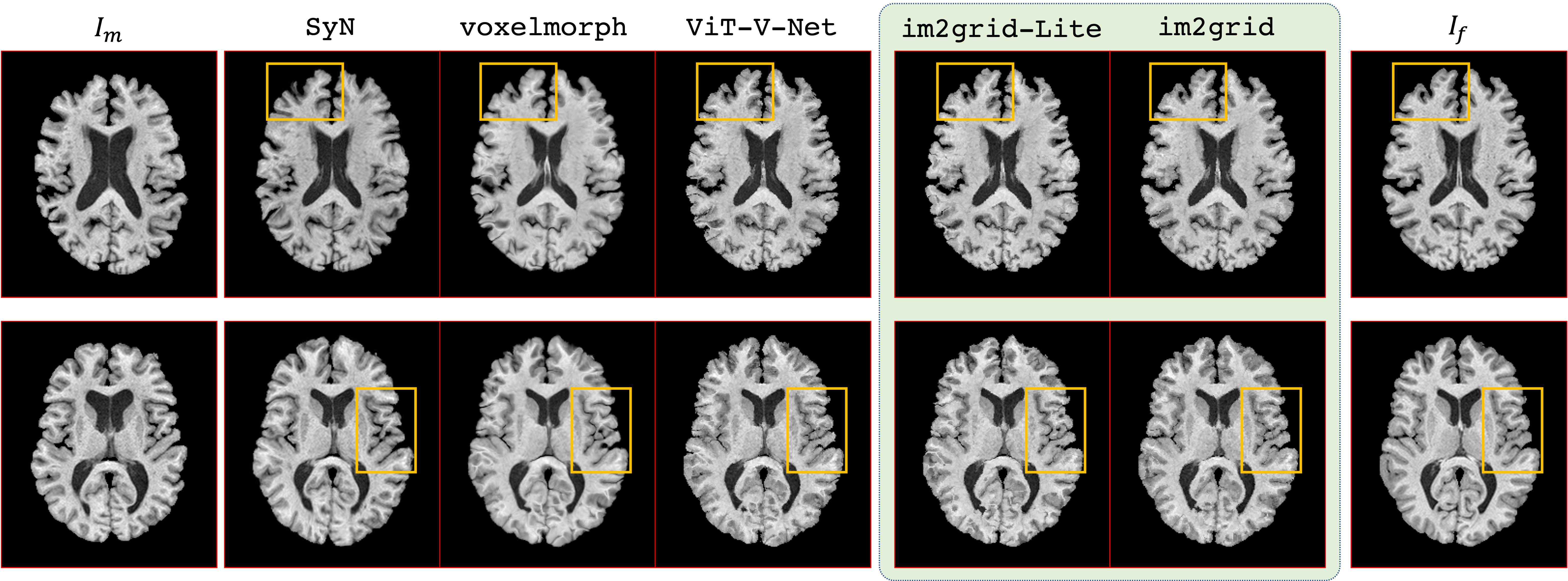}
    \caption{Examples of registering the moving image (the first column) to the fixed image (the last column) using \texttt{SyN}, \texttt{voxelmorph}, \texttt{ViT-V-Net}, and our proposed methods.}
    \label{f:registration_results}
\end{figure}

\subsubsection{Baseline Methods}: We compared our method with several state-of-the-art DL and non-DL registration methods: 1)~\texttt{SyN}: Symmetric image normalization method~\cite{avants2008symmetric}, implemented in the Advanced Normalization Tools~(ANTs)~\cite{avants2009advanced};
2)~\texttt{voxelmorph}: A deep learning based unsupervised method trained with the mean squared error loss~\cite{balakrishnan2019voxelmorph};
3)~\texttt{ViT-V-Net}: A transformer~\cite{vaswani2017attention} based network structure proposed in~\cite{chen2021vit}.

For \texttt{SyN}, a wide range of hyper-parameters were tested on the \texttt{OASIS3} validation set and
the best performing parameters were used for generating the final results.
For \texttt{voxelmorph} and \texttt{ViT-V-Net}, we adopted the same training strategies
as the proposed method, including the loss function and data augmentation.
We optimize the parameters of each method for performance on the \texttt{OASIS3}
validation set and then used those parameters in testing on both datasets.

\subsubsection{Results} For both \texttt{OASIS3} and \texttt{IXI} test datasets, we registered the moving to the fixed image and report the averaged DSC for all labels in Table~\ref{t:results}. In both datasets, the proposed methods outperform the comparison methods for DSC.
For each individual anatomic label, we also conducted a paired,
two-sided Wilcoxon signed rank test
(null hypothesis: the difference between paired
values comes from a distribution with zero median, $\alpha=10^{-3}$) between our methods and the comparison methods. Both proposed methods show significant DSC improvements for seven of nine labels and comparable DSC performance to the best comparison method for the remaining two labels (thalamus and putamen).  
Visual examples on \texttt{OASIS3} data are
shown in Fig.~\ref{f:registration_results}. It can be seen, especially from the highlighted regions, that the warped image produced by the proposed methods have a better agreement with the fixed image.

\begin{table}[!tb]
    \centering
\caption{The Dice coefficient~(DSC), the average number of voxels with negative determinant of Jacobian~(\# of $|J_{\phi}|<0$) and the percentage of voxels with negative determinant of Jacobian~($\%$) for affine transformation, \texttt{SyN}, \texttt{Voxelmorph}, \texttt{ViT-V-Net}, and the proposed methods. The results of the initial alignment by the preprocessing steps are also included. Bold numbers indicate the best DSC for each dataset.}
    \resizebox{1\textwidth}{!}{
    \begin{tabular}{p{0.2\textwidth} C{0.2\textwidth} C{0.2\textwidth} C{0.12\textwidth} C{0.2\textwidth} C{0.2\textwidth} C{0.12\textwidth}}
        \toprule
        & \multicolumn{3}{c}{\textbf{OASIS3}} & \multicolumn{3}{c}{\textbf{IXI}}\\
        \cmidrule(l{2pt}r{2pt}){2-4} \cmidrule(l{2pt}r{2pt}){5-7}
        &  DSC & \# of $|J_{\phi}|<0$ & $\%$ & DSC & \# of $|J_{\phi}|<0$ & $\%$
        \\
        \midrule
        \texttt{Initial} & $0.651\pm0.094$ & $-$ & $0\%$
        & $0.668\pm0.107$ & $-$ & $0\%$
        \\
        \texttt{Affine} & $0.725\pm0.068$ & $-$ & $0\%$
        & $0.748\pm0.052$ & $-$ & $0\%$
        \\
        \texttt{SyN} \cite{avants2008symmetric} & $0.866\pm0.029$ & $223$ & $<0.002\%$
        & $0.845\pm0.035$ & $613$ & $0.008\%$
        \\
        \texttt{Voxelmorph \cite{balakrishnan2019voxelmorph}} & $0.883\pm0.040$ & $85892$ & $<0.7\%$
        & $0.842\pm0.068$ & $21574$ & $<0.3\%$
        \\
        \texttt{ViT-V-Net} \cite{chen2021vit} & $0.872\pm0.042$ & $110128$ & $<0.9\%$
        & $0.845\pm0.068$ & $21298$ & $<0.2\%$
        \\
        \midrule
        \texttt{im2grid-Lite} & $\bm{0.909\pm0.021}$ & $38915$ & $<0.4\%$
        & $\bm{0.870\pm0.043}$ & $14917$ & $<0.2\%$
        \\
        \texttt{im2grid} & $0.908\pm0.023$ & $11880$ & $<0.1\%$
        & $0.865\pm0.050$ & $3235$ & $<0.04\%$
        \\
    \bottomrule
    \end{tabular}
    }
    \label{t:results}
\end{table}

\subsubsection{Evaluation on Learn2Reg Validation Dataset}
We also test the proposed method on the inter-subject brain MRI
registration task from the Learn2Reg
challenge~\cite{hering2021learn2reg}~(L2R 2021 Task $3$).
All scans from the challenge have been preprocessed following~\cite{hoopes2021hypermorph}, and for evaluation purpose segmentation maps of $35$ labels were generated using FreeSurfer~\cite{fischl2012freesurfer}.
We choose the \texttt{im2grid-Lite} version for this task because the challenge evaluation is done on the $\times2$ downsampled images. During training, two scans were randomly selected from the training set and used as input to the proposed method.
The performance is evaluated by comparing the warped segmentation of the moving image and the segmentation of the fixed image. The results are summarized in Table~\ref{t:l2r}, where the DSC represents the average Dice coefficient of all segmented labels; DSC30 is the lowest $30\%$ DSC among all cases, which measures the robustness of the methods; SDlogJ is the standard deviation of the $\log$ of the Jacobian determinant of the deformation field; and HD95 represents the $95\%$ percentile of Hausdorff distance of segmentations. The results of several state-of-the-art methods from the challenge leaderboard are also included. The proposed method shows better accuracy as well as robustness among the comparison methods. Although adopting the instance-specific optimization as described in~\cite{balakrishnan2019voxelmorph} can potential boost the performance on the validation set, our method only used the training set because we assume that such fine tuning process is not available during deployment.

\begin{table}[!t]
    \centering
    \caption{Results of the proposed method and several state-of-the-art methods on the Learn2Reg 2021 Task 3 validation dataset.}
    \resizebox{0.8\textwidth}{!}{
    \begin{tabular}{p{0.22\textwidth} C{0.22\textwidth} C{0.12\textwidth} C{0.12\textwidth} C{0.12\textwidth}}
        \toprule
        & DSC & DSC30 & SDlogJ & HD95 \\
        \cmidrule(l{2pt}r{2pt}){2-5}
        \texttt{im2grid-Lite} & $0.8729\pm0.0142$ & $0.8714$ & $0.1983$ & $1.3786$ \\
        \texttt{TransMorph~\cite{chen2021transmorph}} & $0.8691\pm0.0145$ & $0.8663$ & $0.0945$ & $1.3969$ \\
        \texttt{ConvexAdam~\cite{siebert2021fast}} & $0.8464\pm0.0159$ & $0.8460$ & $0.0668$ & $1.5003$ \\
        Han \etal~\cite{han2021deformable} & $0.8410\pm0.0139$ & $0.8355$ & $0.0796$ & $1.6595$ \\
        Lv \etal~\cite{lv2022joint} & $0.8271\pm0.0131$ & $0.8199$ & $0.1206$ & $1.7220$ \\
        \bottomrule
    \end{tabular}}
    \label{t:l2r}
\end{table}

\section{Discussion}
In this paper, we proposed Coordinate Translator for producing coordinate
correspondences from two feature maps. Additionally, we proposed the \texttt{im2grid}
network that uses Coordinate Translator's for deformable image registration. For unsupervised 3D magnetic
resonance registration, \texttt{im2grid} outperforms the state-of-the-art methods
in accuracy with a similar training and testing speed as other deep learning based
registration methods. Although \texttt{im2grid} has no explicit guarantee of being diffeomorphic, the deformation fields it generated contains fewer voxels with negative determinant of Jacobian compared with other deep learning methods that output deformation fields directly from feature maps. We believe this comes from our design decision to restrict the candidate voxels to the immediate neighborhood of a voxel, which yields a locally smooth deformation field at each scale. We note that even a diffeomorphic algorithm with theoretical guarantees~(\eg SyN) can produce non-diffeomoprhic transformations because of errors introduced during interpolation~\cite{wyburd2021teds}.

For registration, we demonstrated that using Coordinate Translator for matching features and establishing coordinate correspondences together with the convolutional networks for feature extraction can significantly boost the performance. Coordinate Translator is a general module that can be incorporated in
many existing network structures and therefore is not limited
to the registration task. We believe that many tasks that involve image input and coordinate
output can benefit from the use of the Coordinate Translator module.

\subsubsection{Acknowledgement} This work was supported in part by the NIH/NEI grant R01-EY032284 and the Intramural Research Program of the NIH, National Institute on Aging.
%
%
%
\bibliographystyle{splncs04}
\bibliography{citation}
\end{document}

%% file: utils.tex
\usepackage{graphicx}
\usepackage{amssymb}
\usepackage{amsmath}
\usepackage{booktabs}
\usepackage{multirow} 
\usepackage{blindtext}

\usepackage{array}
\newcolumntype{C}[1]{>{\centering\arraybackslash}p{#1}}

\def\eg{\emph{e.g.,\ }}
\def\etal{\emph{et al.}}

\DeclareMathOperator*{\argmin}{arg\,min}

\usepackage{hyperref}
\hypersetup{
    colorlinks=true,
    linkcolor=blue,
    urlcolor=cyan,
    }

%% file: samplepaper.bbl
\begin{thebibliography}{10}
\providecommand{\url}[1]{\texttt{#1}}
\providecommand{\urlprefix}{URL }
\providecommand{\doi}[1]{https://doi.org/#1}

\bibitem{ixi}
{IXI Brain Development Dataset}.
  \url{https://brain-development.org/ixi-dataset/}

\bibitem{avants2008symmetric}
Avants, B.B., Epstein, C.L., Grossman, M., Gee, J.C.: Symmetric diffeomorphic
  image registration with cross-correlation: evaluating automated labeling of
  elderly and neurodegenerative brain. Medical Image Analysis  \textbf{12}(1),
  26--41 (2008)

\bibitem{avants2009advanced}
Avants, B.B., Tustison, N., Song, G., et~al.: Advanced normalization tools
  ({ANTS}). Insight j  \textbf{2}(365),  1--35 (2009)

\bibitem{balakrishnan2019voxelmorph}
Balakrishnan, G., Zhao, A., Sabuncu, M.R., Guttag, J., Dalca, A.V.:
  {VoxelMorph: a learning framework for deformable medical image registration}.
  IEEE Transactions on Medical Imaging  \textbf{38}(8),  1788--1800 (2019)

\bibitem{bazin2007topology}
Bazin, P.L., Pham, D.L.: Topology-preserving tissue classification of magnetic
  resonance brain images. IEEE Transactions on Medical Imaging  \textbf{26}(4),
   487--496 (2007)

\bibitem{cao2017deformable}
Cao, X., Yang, J., Zhang, J., Nie, D., Kim, M., Wang, Q., Shen, D.: {Deformable
  image registration based on similarity-steered CNN regression}. In:
  International Conference on Medical Image Computing and Computer-Assisted
  Intervention. pp. 300--308. Springer (2017)

\bibitem{chen2021transmorph}
Chen, J., Frey, E.C., He, Y., Segars, W.P., Li, Y., Du, Y.: {Transmorph:
  Transformer for unsupervised medical image registration}. arXiv preprint
  arXiv:2111.10480  (2021)

\bibitem{chen2021vit}
Chen, J., He, Y., Frey, E.C., Li, Y., Du, Y.: {V}i{T}-{V}-{N}et: Vision
  transformer for unsupervised volumetric medical image registration. arXiv
  preprint arXiv:2104.06468  (2021)

\bibitem{chou20132d}
Chou, C.R., Frederick, B., Mageras, G., Chang, S., Pizer, S.: {2D/3D image
  registration using regression learning}. Computer Vision and Image
  Understanding  \textbf{117}(9),  1095--1106 (2013)

\bibitem{dalca2018unsupervised}
Dalca, A.V., Balakrishnan, G., Guttag, J., Sabuncu, M.R.: Unsupervised learning
  for fast probabilistic diffeomorphic registration. In: International
  Conference on Medical Image Computing and Computer-Assisted Intervention. pp.
  729--738. Springer (2018)

\bibitem{davatzikos1997spatial}
Davatzikos, C.: {Spatial transformation and registration of brain images using
  elastically deformable models}. Computer Vision and Image Understanding
  \textbf{66}(2),  207--222 (1997)

\bibitem{de2019deep}
De~Vos, B.D., Berendsen, F.F., Viergever, M.A., Sokooti, H., Staring, M.,
  I{\v{s}}gum, I.: A deep learning framework for unsupervised affine and
  deformable image registration. Medical Image Analysis  \textbf{52},  128--143
  (2019)

\bibitem{vos2017end}
{de~Vos}, B.D., Berendsen, F.F., Viergever, M.A., Staring, M., I{\v{s}}gum, I.:
  End-to-end unsupervised deformable image registration with a convolutional
  neural network. In: Deep learning in medical image analysis and multimodal
  learning for clinical decision support, pp. 204--212. Springer (2017)

\bibitem{dosovitskiy2015flownetS}
Dosovitskiy, A., {et al.}: {FlowNet: Learning optical flow with convolutional
  networks}. In: Proceedings of the {IEEE} International Conference on Computer
  Vision. pp. 2758--2766 (2015)

\bibitem{fan2019birnet}
Fan, J., Cao, X., Yap, P.T., Shen, D.: {BIRNet: Brain image registration using
  dual-supervised fully convolutional networks}. Medical Image Analysis
  \textbf{54},  193--206 (2019)

\bibitem{ferrant2000registration}
Ferrant, M., Warfield, S.K., Nabavi, A., Jolesz, F.A., Kikinis, R.:
  {Registration of 3D intraoperative {MR} images of the brain using a finite
  element biomechanical model}. In: International Conference on Medical Image
  Computing and Computer-Assisted Intervention. pp. 19--28. Springer (2000)

\bibitem{fischl2012freesurfer}
Fischl, B.: {FreeSurfer}. NeuroImage  \textbf{62}(2),  774--781 (2012)

\bibitem{fonov-unbiased-2009}
Fonov, V., Evans, A., McKinstry, R., Almli, C., Collins, D.: Unbiased nonlinear
  average age-appropriate brain templates from birth to adulthood. NeuroImage
  \textbf{47}, ~S102 (2009)

\bibitem{gutierrez2016learning}
Guti{\'e}rrez-Becker, B., Mateus, D., Peter, L., Navab, N.: Learning
  optimization updates for multimodal registration. In: International
  Conference on Medical Image Computing and Computer-Assisted Intervention. pp.
  19--27. Springer (2016)

\bibitem{han2021deformableS}
Han, R., {et al.}: {Deformable MR-CT image registration using an unsupervised
  end-to-end synthesis and registration network for endoscopic neurosurgery}.
  In: Medical Imaging 2021. vol. 11598, p. 1159819. International Society for
  Optics and Photonics (2021)

\bibitem{han2021deformable}
Han, R., Jones, C.K., Ketcha, M.D., Wu, P., Vagdargi, P., Uneri, A., Lee, J.,
  Luciano, M., Anderson, W.S., Siewerdsen, J.H.: {Deformable MR-CT image
  registration using an unsupervised end-to-end synthesis and registration
  network for endoscopic neurosurgery}. In: Medical Imaging 2021: Image-Guided
  Procedures, Robotic Interventions, and Modeling. vol. 11598, p. 1159819.
  International Society for Optics and Photonics (2021)

\bibitem{hering2021learn2reg}
Hering, A., Hansen, L., Mok, T.C., Chung, A., Siebert, H., H{\"a}ger, S.,
  Lange, A., Kuckertz, S., Heldmann, S., Shao, W., et~al.: {Learn2Reg:
  comprehensive multi-task medical image registration challenge, dataset and
  evaluation in the era of deep learning}. arXiv preprint arXiv:2112.04489
  (2021)

\bibitem{hoopes2021hypermorph}
Hoopes, A., Hoffmann, M., Fischl, B., Guttag, J., Dalca, A.V.: {Hypermorph:
  Amortized hyperparameter learning for image registration}. In: International
  Conference on Information Processing in Medical Imaging. pp. 3--17. Springer
  (2021)

\bibitem{huo20193d}
Huo, Y., Xu, Z., Xiong, Y., Aboud, K., Parvathaneni, P., Bao, S., Bermudez, C.,
  Resnick, S.M., Cutting, L.E., Landman, B.A.: {3D whole brain segmentation
  using spatially localized atlas network tiles}. NeuroImage  \textbf{194},
  105--119 (2019)

\bibitem{ilg2017flownetS}
Ilg, E., {et al.}: {FlowNet 2.0: Evolution of optical flow estimation with deep
  networks}. In: Proceedings of the {IEEE} Conference on Computer Vision and
  Pattern Recognition. pp. 2462--2470 (2017)

\bibitem{jaderberg2015spatial}
Jaderberg, M., Simonyan, K., Zisserman, A., et~al.: Spatial transformer
  networks. Advances in Neural Information Processing Systems  \textbf{28}
  (2015)

\bibitem{klein2009elastix}
Klein, S., Staring, M., Murphy, K., Viergever, M.A., Pluim, J.P.: Elastix: a
  toolbox for intensity-based medical image registration. IEEE Transactions on
  Medical Imaging  \textbf{29}(1),  196--205 (2009)

\bibitem{lamontagne2019oasis}
LaMontagne, P.J., Benzinger, T.L., Morris, J.C., Keefe, S., Hornbeck, R.,
  Xiong, C., Grant, E., Hassenstab, J., Moulder, K., Vlassenko, A.G., et~al.:
  {OASIS-3: Longitudinal neuroimaging, clinical, and cognitive dataset for
  normal aging and Alzheimer disease}. MedRxiv  (2019)

\bibitem{lv2022joint}
Lv, J., Wang, Z., Shi, H., Zhang, H., Wang, S., Wang, Y., Li, Q.: {Joint
  progressive and coarse-to-fine registration of brain MRI via deformation
  field integration and non-rigid feature fusion}. {IEEE} Transactions on
  Medical Imaging  (2022)

\bibitem{reinhold2019evaluatingS}
Reinhold, J.C., {et al.}: {Evaluating the impact of intensity normalization on
  MR image synthesis}. In: Medical Imaging 2019: Image Processing. vol. 10949,
  p. 109493H. International Society for Optics and Photonics (2019)

\bibitem{ronneberger2015u}
Ronneberger, O., Fischer, P., Brox, T.: {U-Net: Convolutional networks for
  biomedical image segmentation}. In: International Conference on Medical image
  computing and computer-assisted intervention. pp. 234--241. Springer (2015)

\bibitem{rueckert1999nonrigid}
Rueckert, D., Sonoda, L.I., Hayes, C., Hill, D.L., Leach, M.O., Hawkes, D.J.:
  {Nonrigid registration using free-form deformations: application to breast MR
  images}. IEEE Transactions on Medical Imaging  \textbf{18}(8),  712--721
  (1999)

\bibitem{siebert2021fast}
Siebert, H., Hansen, L., Heinrich, M.P.: {Fast 3D registration with accurate
  optimisation and little learning for Learn2Reg 2021}. In: International
  Conference on Medical Image Computing and Computer-Assisted Intervention. pp.
  174--179. Springer (2021)

\bibitem{thirion1998image}
Thirion, J.P.: {Image matching as a diffusion process: an analogy with
  Maxwell's demons}. Medical Image Analysis  \textbf{2}(3),  243--260 (1998)

\bibitem{tustison2010n4itk}
Tustison, N.J., Avants, B.B., Cook, P.A., Zheng, Y., Egan, A., Yushkevich,
  P.A., Gee, J.C.: {N4ITK: improved N3 bias correction}. IEEE Transactions on
  Medical Imaging  \textbf{29}(6),  1310--1320 (2010)

\bibitem{vaswani2017attention}
Vaswani, A., Shazeer, N., Parmar, N., Uszkoreit, J., Jones, L., Gomez, A.N.,
  Kaiser, {\L}., Polosukhin, I.: Attention is all you need. Advances in Neural
  Information Processing Systems  \textbf{30} (2017)

\bibitem{vercauteren2009diffeomorphic}
Vercauteren, T., Pennec, X., Perchant, A., Ayache, N.: {Diffeomorphic demons:
  Efficient non-parametric image registration}. Neuro{I}mage  \textbf{45}(1),
  S61--S72 (2009)

\bibitem{wyburd2021teds}
Wyburd, M.K., Dinsdale, N.K., Namburete, A.I., Jenkinson, M.: {TEDS-Net:
  Enforcing diffeomorphisms in spatial transformers to guarantee topology
  preservation in segmentations}. In: International Conference on Medical Image
  Computing and Computer-Assisted Intervention. pp. 250--260. Springer (2021)

\bibitem{yang2017quicksilver}
Yang, X., Kwitt, R., Styner, M., Niethammer, M.: {Quicksilver: Fast predictive
  image registration--a deep learning approach}. NeuroImage  \textbf{158},
  378--396 (2017)

\end{thebibliography}
